\documentclass[reprint,
 amsmath,amssymb,
 aps,
]{revtex4-2}

\usepackage{graphicx}
\usepackage{dcolumn}
\usepackage{bm}

\begin{document}

\preprint{APS/123-QED}

\title{Stabilization of Stone-Wales Defects in Metal-supported Graphene }

\author{Rob H. Mason}
 \altaffiliation{School of Physics and Applied Physics, Southern Illinois University Carbondale , IL 62901, USA}
\author{Manuka M. S. Sinharage}
 \altaffiliation{School of Physics and Applied Physics, Southern Illinois University Carbondale, IL 62901, USA}
 \author{Hansika I. Sirikumara}
 \altaffiliation{E. S. Witchger School of Engineering, Marian University, Indianapolis, IN 46222, USA}
 \author{Sabrina Nilufar}
 \altaffiliation{School of Mechanical Engineering, Southern Illinois University Carbondale , IL 62901, USA}
 \email{sabrina.nilufar@siu.edu}
 \author{Thushari Jayasekera}
 \altaffiliation{School of Physics and Applied Physics, Southern Illinois University Carbondale , IL 62901, USA}
 \email{thushari@siu.edu}




\date{\today}

\begin{abstract}

The characteristics of graphene–metal interfaces play a decisive role in their electronic, optoelectronic, and mechanical applications. Properties such as charge transfer across the interface become particularly significant in the presence of topological defects. The stability of Stone-Wales (SW) defects in graphene is governed by the balance between three energy descriptors, the activation energy, formation energy, and restoration energy. By comparing the energy parameters obtained from first-principles density functional theory calculations, we show that SW defect formation is energetically more favorable on metal-supported graphene. Our calculations for SW defects in graphene/Cu(111) and graphene/Al(111) systems indicate only a little dependence of energy profile on the type of metal. The presence of the metal substrate leads to a $\sim$ 12\% increase in the formation energy and a $\sim$20\% reduction in the activation energy, which together favor the formation of Stone–Wales defects. Although the restoration energy decreases by about $\sim$35\% in metal-supported graphene, it remains significantly higher to prevent self-healing. As a result, once formed, the Stone–Wales defects are likely to remain stable, suggesting the possibility of terminal SW defect formation in metal-supported graphene. 
\end{abstract}

\maketitle


\section{INTRODUCTION\label{Intro}}
Metal matrix composites (MMC) are highly desirable for applications that require strong yet lightweight materials for use in extreme conditions, such as electrical transmission systems and building infrastructure. 
The formation of different bonds at the metal/graphene interface can cause agglomeration and phase separation of the metal matrix making the synthesis of graphene-reinforced MMCs challenging \cite{ju2017facile,sadhu2023sic,guler2020short}. The formation of defects in graphene resulting from metal-carbon interactions is reported as a crucial factor for graphene to be an effective reinforced material for MMCs \cite{chen2020advances,banhart2009interactions,cao2023review}.  
Using the formation energy calculations and the charge transfer analysis within the first principles Density Functional Theory (DFT), we demonstrated that the interfacial interaction at the metal/graphene interfaces promotes the formation and stabilization of Stone-Wales defects.

Graphene is a 2-dimensional network of $sp^2$-hybridized carbon, which has been widely studied for both electrical and mechanical applications for its high strength and high electron mobility. 
While it is not possible to synthesize defect-free graphene, fully pristine graphene is not suitable for many applications. Stone–Wales (SW) defects are a fundamental type of topological defect in $sp^2$-hybridized  materials such as graphene.

Graphene with controlled SW defects has gained significant interest due to its ability to enhance the strength of metal matrix composites (MMCs) as well as its potential in various other applications, including gas sensors, energy storage, and more   \cite{jovanovic2023reactivity,liu2014improving,er2016defective,li2017effect,SiGrSi}. SW defects could be either beneficial or detrimental depending on the target property. A deeper understanding of its formation at various interfaces is essential for achieving controlled SW defect growth.

An SW defect can be considered as being formed by an in-plane $90^0$ rotation of a C-C bond, converting four adjacent hexagons into two pentagons and two heptagons; therefore, it is commonly referred to as a 5-7-7-5 defect \cite{stone1986theoretical,PRB_LiReich}. While the Stone--Wales defect is often described as an in-plane topological defect generated by bond rotation within the carbon lattice, several studies have suggested that additional wave-like out-of-plane distortion may contribute to its stabilization.
\cite{ertekin2009topological,yazyev2010topological,banhart2011structural}.

\begin{figure}[!b]
    \centering
     \includegraphics[width=0.95\linewidth]{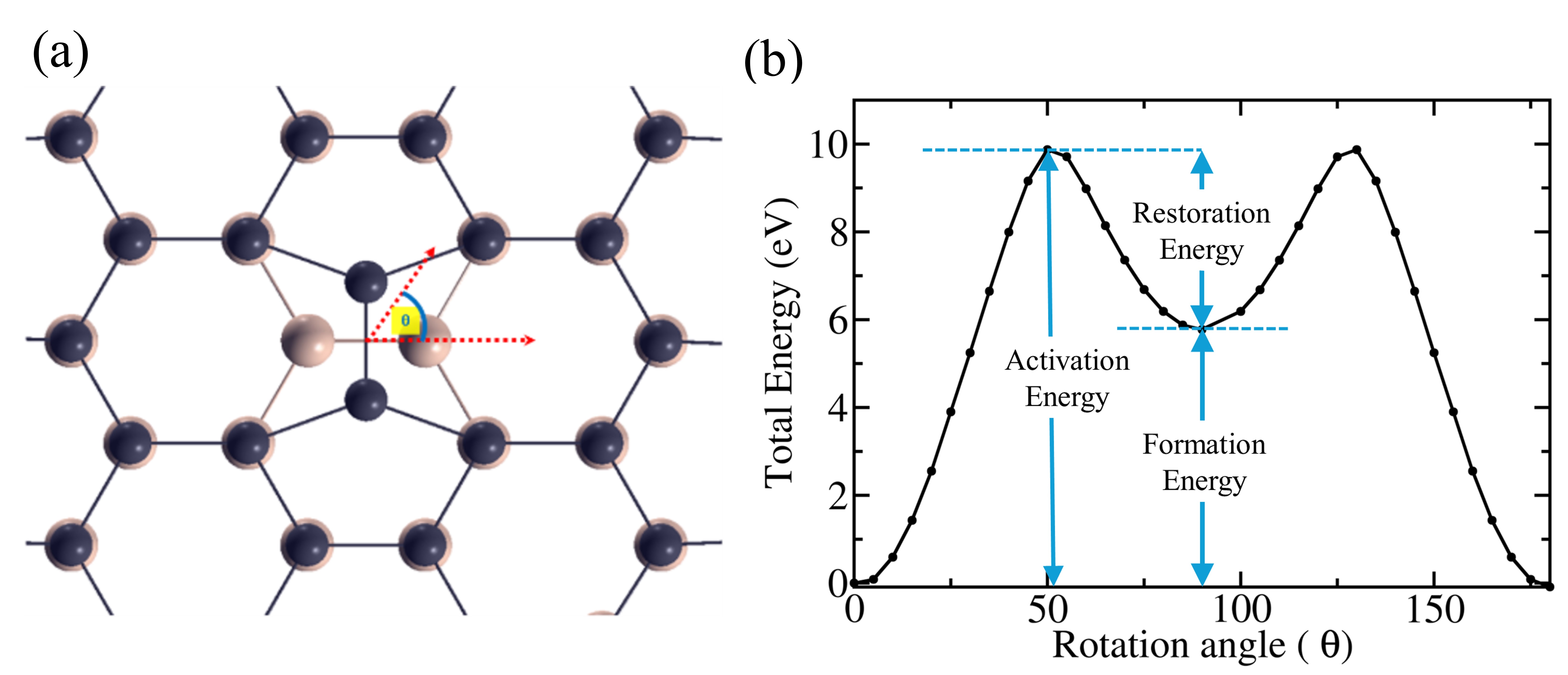}
    \caption{ (a) Top view of the graphene lattice showing the c-c dimer selected for rotation to evaluate the energy profile associated with Stone–Wales (SW) defect formation. The rotation angle $\theta$ of the c-c dimer is indicated (b) Total energy as a function of the dimer rotation angle $\theta$. The activation, formation, and restoration energies are indicated. }
    \label{PrGr}
\end{figure}

The stability of SW defects can be characterized by three key energy descriptors: activation energy ($E_a)$, formation energy ($E_f$), and restoration energy ($E_r$); which can be determined from the density functional theory total energy profile constructed as a function of the C–C bond rotation angle, $\theta$\ (Fig \ref{PrGr}) \cite{PRB_LiReich}. In this context, pristine graphene corresponds to $\theta =0$ and $\theta=90^o$  represents the SW defect state. The activation energy is the energy difference between pristine graphene and the transition state. The formation energy refers to the energy difference between graphene with SW defects and pristine graphene, while the restoration energy is the energy difference between graphene with SW defects and the transition state.
The calculated formation energy of the Stone–Wales (SW) defect is 5.8 eV (Fig \ref{PrGr}), consistent with reported values of 4.6–5.8 eV \cite{PRB_LiReich}. The activation barrier is 9.9 eV, while the restoration energy is 4.1 eV, all in agreement with previously reported ranges \cite{ma2009stone,kotakoski2011stone,kotakoski2006energetics,krasheninnikov2010ion}. Due to the higher restoration energy, if SW defects are formed, their self healing to the pristine structure is highly unlikely, and the SW defects become terminal defects \cite{bhatt2022various}. In fact, SW defects formation of graphene is reported via high temperature and irradiation \cite{kotakoski2011stone}.

It has been reported that the C–C bonds associated with Stone–Wales (SW) defects are more reactive than those in pristine graphene \cite{jovanovic2023reactivity}. Metal adatoms preferentially stabilize at the hollow sites of SW defects compared to pristine graphene \cite{harman2014density}. While SW defects can be advantageous for certain applications, interactions of adatoms with these defects have also been investigated in the context of restoring defect-free graphene. W. W. Wang et al. reported that transition-metal adatoms can reduce the SW restoration barrier to 2.9 eV \cite{wang2012metal}, while C. Wang et al. showed that adsorption of a carbon adatom further lowers the restoration energy to 0.86 eV \cite{wang2013catalytically}. Previous studies report that the activation barrier for Stone–Wales defect formation in silicene is reduced from approximately 2.4 eV in free-standing silicene to about 1.6 eV when supported on Ag(111) \cite{silecene}.

B.P Klein et al reported that SW defects induce bonding and charge transfer on Cu substrate \cite{klein2022topological,D3NR04690G}. By comparing azupyrene (with a 5-7-7-5 rings) to pyrene (with 6-member rings)  as model systems, B.P. Klein et al. demonstrated that SW defects exhibit stronger binding on the Cu(111) surface. Azupyrene shows a localized
electron transfer from the metal to the graphene sheet, which result in a reduced metal-graphene distance. Based on density functional theory calculations for graphene on Cu(111) containing a single 5–7–7–5 defect site, B. P. Klien concluded that graphene with an SW defect binds more strongly to the Cu(111) surface than pristine graphene \cite{klein2022topological}.

This article reports on the Stone–Wales (SW) defect formation energy profile of graphene on metal substrates, using Cu(111) and Al(111) as representative systems. Our results suggest that charge transfer  at the metal/graphene interface significantly modifies the energy profile of SW defect formation, thereby increasing the likelihood of defect formation in metal-supported graphene.

\section{Calculation Details\label{sec:level1calc}}

\begin{figure}[t]
    \centering
    \includegraphics[width=1 \linewidth]{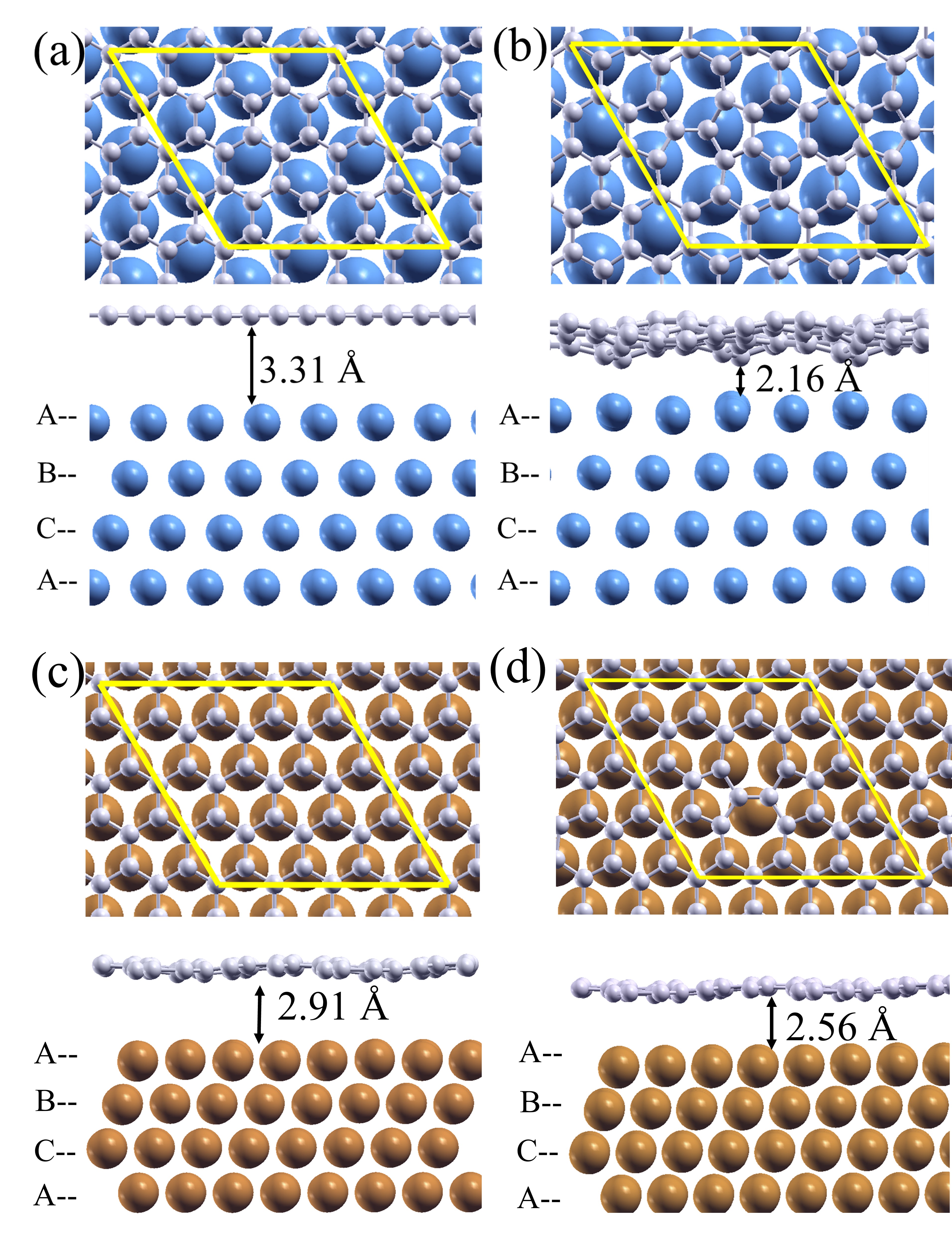}
    \caption{Top and side views of the atomic configurations of the metal–graphene interfaces used in this study. (a) Graphene and (b) graphene with a Stone–Wales (SW) defect on Al(111), and (c) graphene and (d) graphene with an SW defect on Cu(111). In each panel, the top row shows the top view, while the bottom row shows the side view. The closest metal–graphene distance is indicated. The yellow parallelepiped represents the supercell used in the DFT calculations. }
    \label{topsideview}
\end{figure}

We investigated the energy profile for SW defect formation in  graphene supported by Al(111) and Cu(111) surfaces.  Fig \ref{topsideview} shows the top and side views of the atomic configurations of pristine and Stone–Wales-defective graphene on Al(111) and Cu(111) surfaces, as used in this work.
For aluminum, the optimized bulk lattice constant is 4.05 \AA, giving an in-plane lattice constant of 2.86 \AA  for the Al(111) surface, approximately 16\% larger than that of graphene (2.46 \AA) \cite{Reba}. Therefore, a direct match of 1×1 would induce excessive strain. Instead, a $\sqrt{3} \times \sqrt{3} R \, \text{cos} \, 30^o$ Al(111) (R3) supercell aligned with a 2×2 graphene cell reduces the mismatch to $\sim$0.8\%, providing a nearly commensurate interface suitable for periodic calculations such as Density Functional Theory. In fact, this configuration was used in previous calculations of Al/Gr interfaces \cite{AlR3PRL,AlR3PRB}. We considered 4x4 graphene on 2x2 R3 cells of Al(111) with 32 C atoms and 12 Al atoms at the interface for our calculations. For Cu(111), with a surface lattice constant of 2.56 \AA, a 1×1 graphene cell introduces only $\sim 4$\% strain. We therefore used a 1×1 Cu(111)/1×1 graphene interface, corresponding to 16 Cu atoms and 32 C atoms at the interface. The graphene layer was strained to match the underlying metal substrate, an approach that has been reported as realistic in previous studies \cite{adamska2012atomic}.   For both Al(111) and Cu(111), we considered two configurations: graphene supported on the metal substrate and graphene sandwiched within the metal matrix. The sandwithced configurations mimic the graphene-reinforced metal-matrix configurations. 
For the case of metal-supported graphene (Al/Gr and Cu/Gr), at least 12 \AA\ vacuum space was used to prevent spurious interactions between periodic images. For Metal/Graphene/Metal systems (Al/Gr/Al and Cu/Gr/Cu), the cell dimension perpendicular to the graphene sheet is optimized to achieve the minimum energy
configurations with the graphene in its pristine state. An SW defect is generated by a $90 ^o$ rotation of a C–C dimer within the graphene plane about an axis passing through the bond and normal to the plane.

Density functional theory calculations were performed using the plane-wave pseudopotential method as implemented in the Quantum Espresso package \cite{QE}. Exchange–correlation effects were treated within the generalized gradient approximation (GGA) including van der Waals corrections. Previous work by Adamska et al. reported that results obtained for metal–graphene interfaces using GGA are essentially identical to those obtained using the local density approximation (LDA) \cite{adamska2012atomic}.
The plane-wave kinetic-energy cutoff was set to 40 Ry for the wave functions. Brillouin-zone integrations were performed using a Monkhorst–Pack 
k-point mesh of 6 × 6 × 1. Atomic positions were relaxed until the forces on each atom were less than 0.01 eV/\AA, while the total energy convergence threshold was set to $10^{-6}$ eV.

 \section{Results\label{sec:level1Res}}

\begin{figure}[!h]
    \centering
        \includegraphics[width=0.95\linewidth]{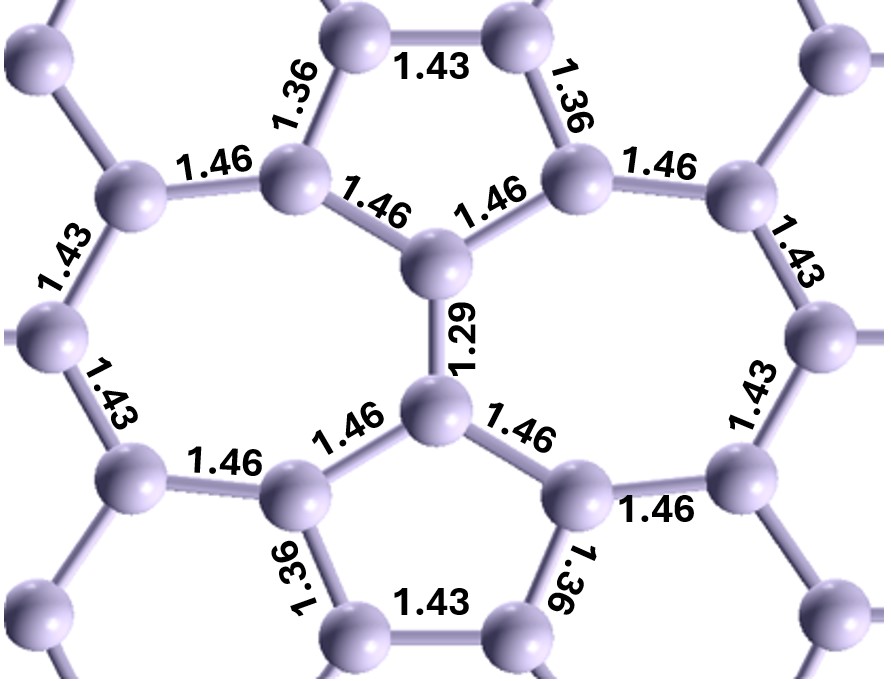}
     \includegraphics[width=0.95\linewidth]{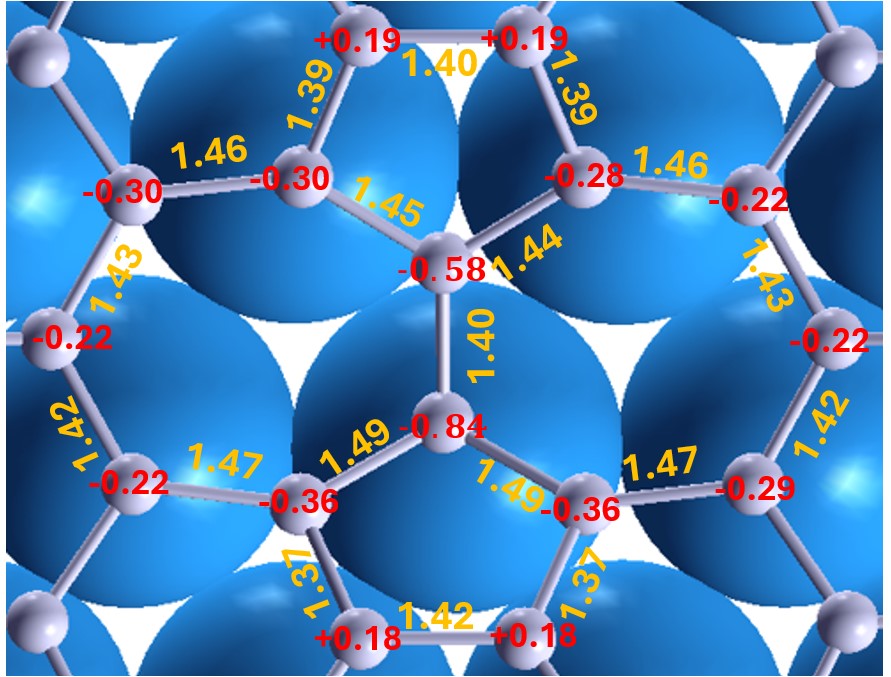}
        \includegraphics[width=0.95\linewidth]{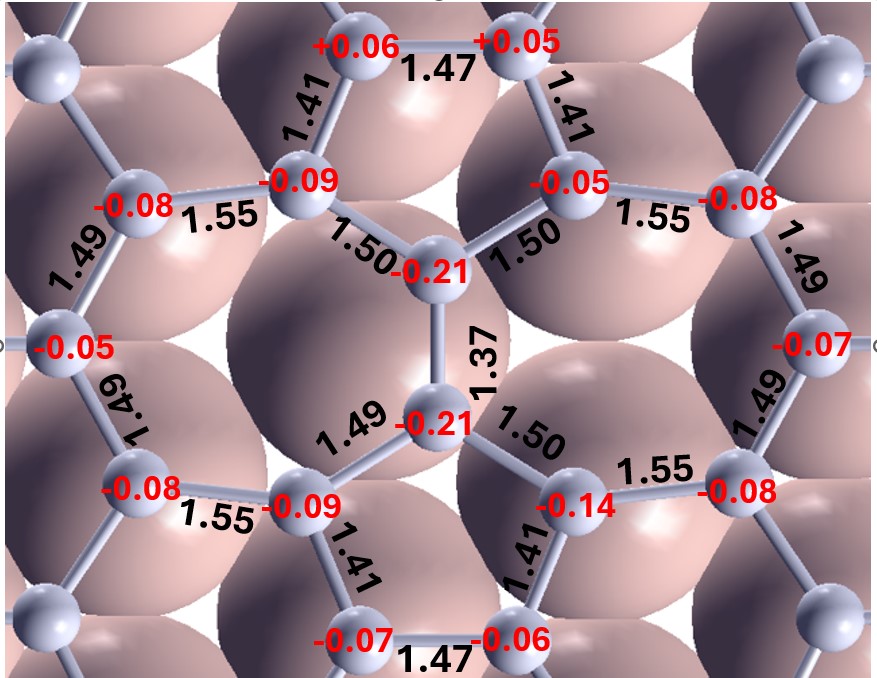}
    \caption{Bond-length distribution around the Stone-Wales defect for free-standing graphene and for graphene on Cu(111) and Al(111) surfaces. All bond lengths (marked along the bonds) are given in \AA, while the values shown on the atoms indicate the vertical displacement (in \AA) of each atom relative to its average position. Carbon, Alumnium and Copper atoms are shown in grey, blue and metallic color spheres. }
    \label{SW_bondlength}
\end{figure}
Fig \ref{SW_bondlength} shows the bond-length distribution around the Stone–Wales defect for free-standing graphene and for graphene on the Cu(111) and Al(111) surfaces. All bond lengths (marked along the bonds) are given in \AA, while the values shown on the atoms indicate the vertical displacement (in \AA) of each atom relative to its average position.
As reported in the literature, the C–C bond length decreases to $\sim $ 1.28 \AA \,  for the Stone–Wales defect in free-standing graphene. In contrast, when graphene is supported on Al and Cu surfaces, the C–C bond length decreases only to $\sim$ 1.40 \AA (Fig. \ref{SW_bondlength}). This difference indicates that interactions at the metal-graphene interface influences the structural details of the Stone–Wales defect. The geometric and energetic parameters of the three systems are summarized in Table \ref{TableData}. Additionally, the length ($L_{\text{SW}}$), width ($W_{\text{SW}}$), and height , i.e. ripples ($R_{\text{SW}}$) of the 18-atom defect site are noted in Table.\ref{TableData}.

To investigate the energy profile associated with the SW defect formation, a selected C-C bond  was rotated stepwise from $\theta=0^o$ to $180^o$. Here $\theta=0^o$ and $180^o$ represent pristine graphene, while $\theta=90^o$ represents the SW defective graphene configuration. During these steps, the direction of the C-C bond was constrained to fix $\theta $ at specific angles. For free-standing graphene, the C–C bond length at the defect site was adjusted to achieve the lowest-energy state for each $\theta$ value. In contrast, for metal-supported graphene, C-C bond length at the defect site was kept fixed through out the rotation, as bond shrinkage is much less pronounced compared to the bond shrinkage in free-standing graphene (Fig \ref{SW_bondlength}).

\begin{table}[h]
\label{tab:example}
\begin{ruledtabular}
\begin{tabular}{ccccccccc}
System & $E_b$ & $E_r$ & $E_f$ & $d_s$ & $W_{SW}$ & $L_{SW}$& $d_{C-C}$ &$R_{SW}$\\
FS-Gr & 9.9 & 4.1 & 5.8 & - & 5.31 & 6.80 & 1.29 & 0.01 \\
AlGr        & 7.65  & 2.55  &  5.10    &   2.16   &  5.18  & 6.74  & 1.40 & 1.48\\
AlGrAl      & 8.42 & 3.62 & 4.80     &   2.16   &  5.18    &  6.74  & 1.40 & 1.07\\
CuGr        & 7.71  & 2.36  &  4.35    &  2.56    &  5.54    &  7.06 & 1.37 & 0.40\\
CuGrCu      & 7.73 & 2.34 &   4.39   &  2.56    & 5.54     &  7.05 &1.36 & 0.04 \\ 
\end{tabular}
\end{ruledtabular}\caption{The energy  and the structural parameters of Gr/Metal systems in comparison with SW defects in free standing graphene (FS-Gr). $E_a$ -Activation energy, $E_f$ - Formation energy, $E_r$ Restoration Energy, $d_s$ - the shortest Metal-Gr Distance. $W_{\text{SW}}$ - Width of SW site, $L_{\text{SW}}$ - the length of SW site, $R_{\text{SW}}$ - the ripple width of graphene layer, and $d_{C-C}$ bond length of SW defect site, }
\label{TableData}
\end{table}

SW formation energy is calculated as
$$E_f=E_{{\text{system}}}-E_{{\text{metal/graphene}}},$$
where $E_{{\text{system}}}$ is the total energy of the fully optimized SW-graphene/metal configuration, $E_{{\text{metal/graphene}}}$ is the same system optimized with graphene in its honeycomb network with no bond rotation.

\begin{figure}[!b]
    \centering
     \includegraphics[width=1 \linewidth]{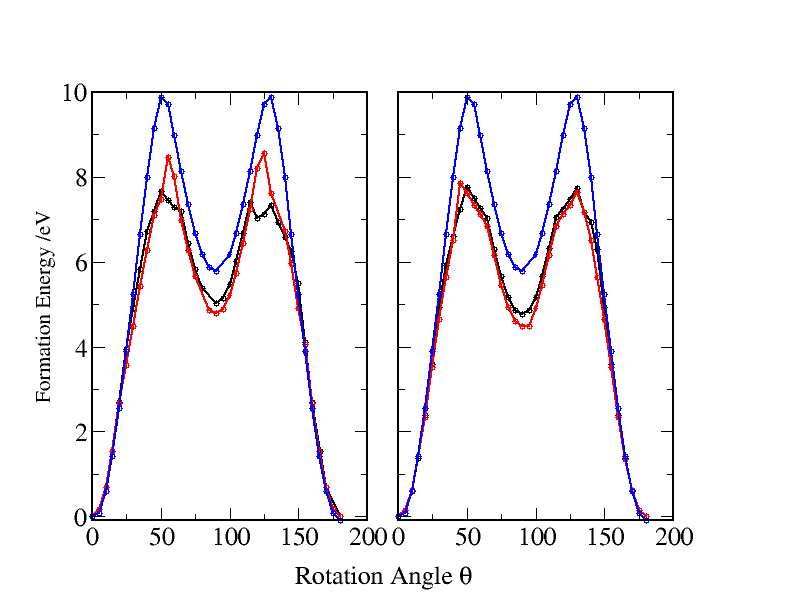}
    \caption{Energy profile of SW defect formation: $E_f$ is plotted as a function of the dimer rotation angle for graphene on Al(111) in the left and on Cu(111) in the right panel. Energy profiles for FS-Gr is shown in blue, metal/Gr in black and metal/Gr/metal in red.  }
    \label{figEf}
\end{figure}

Fig \ref{figEf} shows the energy profile for the SW defect formation on Al(111) and Cu(111) as a function of dimer rotation  angle. Transition state of SW formation on FS-Gr was found to be ~$45^0-50^0$  with an activation energy barrier of ~$9.9$ eV  (also shown in Fig \ref{PrGr}) is shown here for comparison \cite{PRB_LiReich}.

\begin{figure}[!b]
    \centering
     \includegraphics[width=0.9\linewidth]{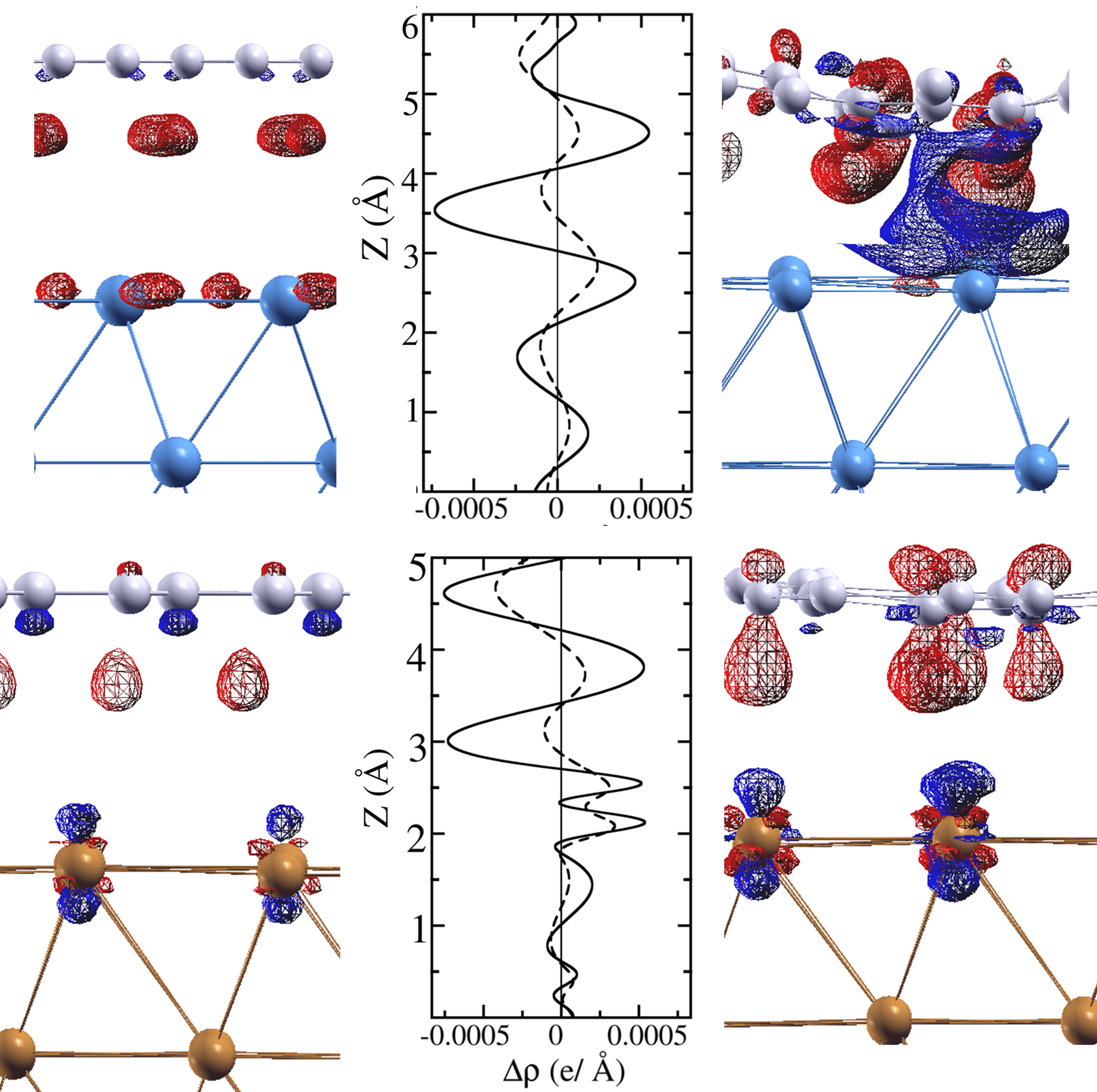}
    \caption{Charge transfer for pristine graphene and graphene with a Stone–Wales (SW) defect on (a) Al(111) and (b) Cu(111). The left and right panels show charge-transfer density isosurfaces for pristine and SW-defected graphene, respectively (red: charge depletion; blue: charge accumulation). The center panel shows the planar-averaged charge density, with dashed and solid lines representing pristine and SW-defected graphene, respectively.}
    \label{figdeltarho}
\end{figure}

\begin{figure}[!h]
    \centering
     \includegraphics[width=0.97\linewidth]{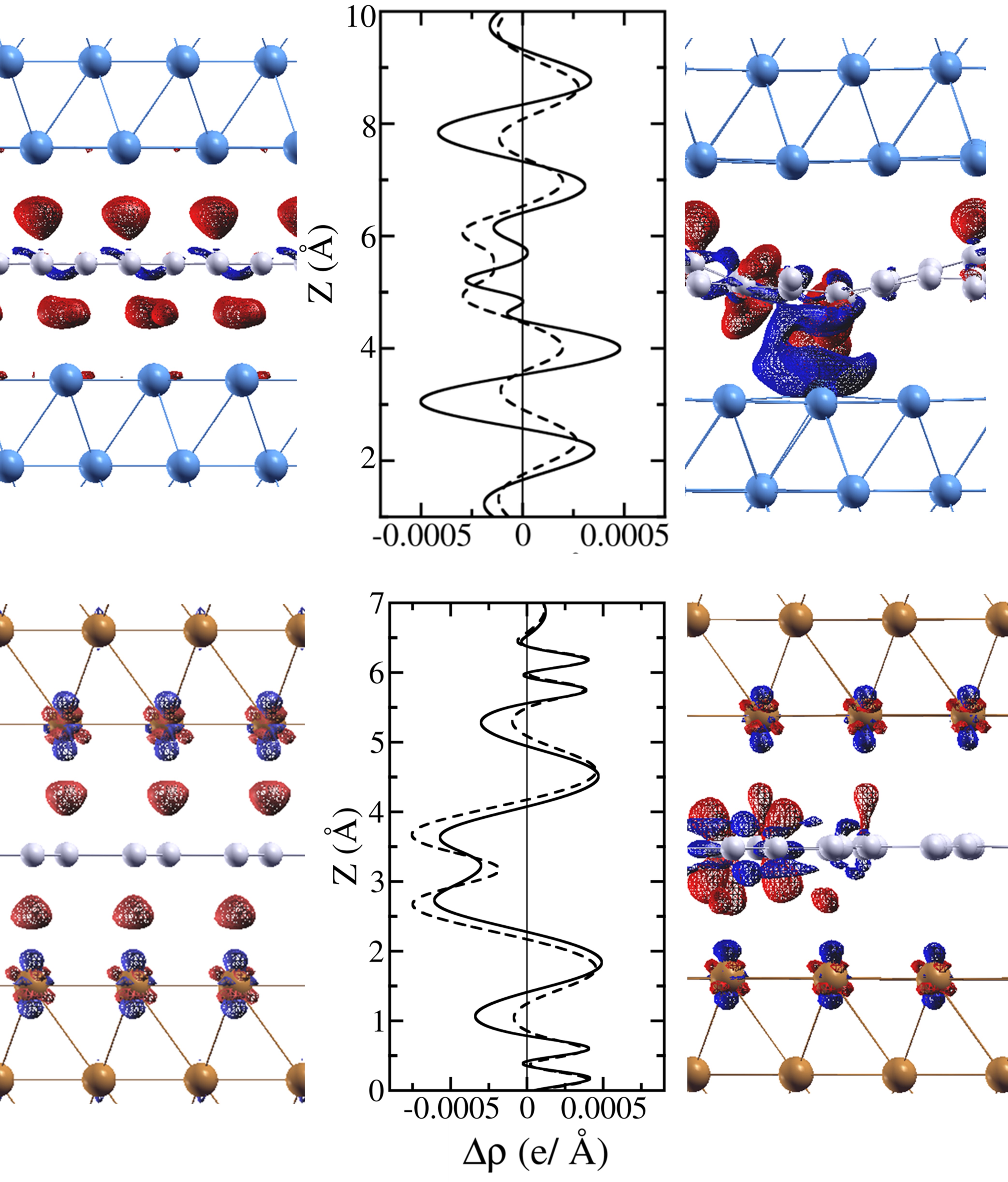}
    \caption{Charge transfer for pristine graphene and graphene with a Stone–Wales (SW) defect on (a) Al(111) and (b) Cu(111) in sandwich configurations. The left and right panels show charge-transfer density isosurfaces for pristine and SW-defected graphene, respectively (red: charge depletion; blue: charge accumulation). The center panel shows the planar-averaged charge density, with dashed and solid lines representing pristine and SW-defected graphene, respectively. }
    \label{figdeltarho_sandwitched}
\end{figure}

The activation energy decreases for metal-supported graphene; for example, graphene on Al(111) (AlGr) exhibits an activation energy of 7.65 eV, corresponding to an $\sim$ 22\% reduction compared to that on free-standing graphene. A comparable reduction is observed for the other three systems. In addition, the SW defect in graphene on Al is $\sim$ 12\% more stable than in free-standing graphene. These results suggest a higher likelihood of SW defect formation in metal-supported graphene. We also observe a reduction of $\sim$ 35\% in the restoration energy of SW defects on Al-supported graphene.
Although the restoration barrier is reduced, it is still  sufficiently high to prevent spontaneous self healing. Once SW defects are generated through high-energy experimental processes, such as irradiation, these topological defects can still become terminal defects. 
Our results indicate only a weak dependence of the energy profile on the type of metal substrate (Fig. \ref{figEf}).  We examined the charge transfer in pristine and SW-defected graphene at Al(111) and Cu(111) interfaces for both sandwiched and on-top configurations.

Charge transfer is calculated by 
$$\Delta \rho = \rho_{{\text{system}}}-\rho_{{\text{Metal}}}-\rho_{{\text{graphene}}}$$
In the case of charge transfer calculations, the charge density of the metal substrate ($\rho_{{\text{meta}}}$ ) and graphene layer ($\rho_{{\text{graphene}}}$) is calculated for the exact same atomic configuration without geometry optimizations.

Fig \ref{figdeltarho} illustrates the charge transfer characteristics of pristine graphene and graphene with SW defects on Al(111) and Cu(111) substrates. In the pristine systems, graphene on both Al(111) and Cu(111) exhibits negligible charge transfer to the substrate as shown in the left panels of the Fig \ref{figdeltarho}. SW defects enhances the graphene/metal interactions, which result in ripples in graphene  This is caused by the increase in charge transfer as shown in the right panel of the Fig \ref{figdeltarho}. The central panel of Fig. \ref{figdeltarho}) shows the planar average charge density. 

Fig \ref{figdeltarho_sandwitched} shows the charge transfer in pristine and SW-defected graphene at Al/Gr/Al and Cu/Gr/Cu systems.  In both systems, charge accumulates around the graphene layers, while charge depletion occurs at the substrate, which is consistent with the on-top-configuration results.

\section{Conclusion}

DFT total-energy calculations suggest that SW defect formation is energetically more favorable in metal-supported graphene. A similar tendency is observed in graphene-reinforced metal matrix configurations, which was mimicked by metal/Gr/metal sadnwiched configurations. This effect is likely associated with enhanced charge transfer between SW-defective graphene and the metal substrates. Such understanding may help guide the controlled utilization of SW defects in metal matrix composites designed for extreme environments.

\section{Acknowledgment}

This work used  Anvil at Purdue University and Stampede3 at the University of Texas through allocation BIO240065, PHY250285, and MAT250109 from the Advanced Cyberinfrastructure Coordination Ecosystem: Service and Support (ACCESS) program which is supported by U.S. National Science Foundation grants $\#2138259, \#2138286, \#2138307, \#2137603, and \#2138296$.
We also used the High-Performance Computing Cluster at Southern Illinois University Carbondale.
RM acknowledges the Dissertation Research Assistantship Award through  Southern Illinois University Carbondale.

\bibliography{RMason}

\end{document}